# Corporate Governance, Noise Trading and Liquidity of Stocks

Jianhao Su[*]

## Abstract

Our main task is to study the effect of corporate governance on the market liquidity of listed companies' stocks. We establish a theoretical model that contains the heterogeneity of investors' beliefs to explain the mechanisms by which corporate governance improves liquidity of the corporate stocks. In this process we found that the existence of noise traders who are semi-informed in the market is an important condition for corporate governance to have the effect of improving liquidity of the stocks. We further find that the strength of this effect is affected by the degree of noise traders' participation in market transactions. Our model reveals that corporate governance and the degree of noise traders' participation in transactions have a synergistic effect on improving the liquidity of the stocks.

## 1. Introduction

Shleifer and Vishny (1997) define that the purpose of corporate governance is a series of constraints and institutional arrangements formulated to solve the two types of agency conflicts and protect the interests of investors.

The liquidity of stocks studied in most of the existing literature refers to the liquidity of stocks in the secondary market. The liquidity that we focus on in this paper is also the liquidity of stocks in the secondary market. According to Harris (1990), a security is said to have higher liquidity if it can be sold in large quantities at lower transaction costs in a short period of time and its market price is less affected. The level of stock market liquidity determines the level of utility investors get in their investments, so it also affects their investment decisions. Therefore, Securities liquidity has been a research hotspot in the financial market field, and a large number of related classic literatures have emerged, such as: Amihud (1986, 2002), Pastor and Stambaugh (2003), Acharya and Pedersen (2005), Liu (2006), Amihud and Hameed et al. (2015), etc. If an investor holds a security that has lower liquidity, he has to sell it at a price significantly lower than the true value of the security when for some reason he needs to sell it in large quantities within a short period of time, as will bring him a greater loss. For securities with better liquidity, the holder can sell a larger number of such securities at a price close to its fair value in a short period of time, without causing a significant decline in the price of such securities. When the prospects of the securities market are not optimistic and investors are not confident, a large number of securities will be sold off, which will cause the prices of various securities in the market to decrease. When the prospects of the securities market are not optimistic and investors are not confident, a large number of securities will be sold off, which will cause the prices of various securities in the market to fall. But at this time, the prices of securities with higher liquidity tend to have a slighter decrease, so such securities play a role in stabilizing the market in decline.

Since 2002, more and more literature has begun to focus on the relationship between corporate governance and micro characteristics that the corporate stocks exhibit in market transactions. Gompers and Ishii et al. (2003), Aggarwal et al. (2009), Bebchuk and Cohen et al. (2009), Johnson, Moorman et al. (2011), etc. find that the return on the stocks of companies with better protection of investor rights has a higher premium and the co mpany's market value is also higher. The main topic of this paper is to study the impact of corporate governance on the market liquidity of listed companies' stocks. Using empirical tests, Bacidore, Sofianos (2002) and Chung (2006) show that stocks of the companies in a more stringent external governance environment exhibit higher liquidity. Chen and Chung (2007), Chung and Elder et al. (2010), Li et al. (2012), Prommin et al. (2014), Tang and Wang (2015) use the data about listed companies in different countries to conduct their empirical examinations respectively, but they get an almost consistent conclusion that there is a significant positive correlation between the quality of a company's internal governance and the level of the liquidity of its stocks. Existing literature focuses on seeking empirical evidence that there is a correlation between corporate governance and liquidity of corporate stocks, while in this paper we mainly analyze the conditions and mechanisms by which corporate governance affects the liquidity of corporate stocks. The issue studied in this paper lies in the intersection of the two fields of securities investment and corporate governance.

---

[*] Jianhao Su, 201720030@mail.sdu.edu.cn, school of economics, Shandong University.

In the analysis of this paper, a company's executives and its controlling shareholders who can control the company in a large degree are considered as a whole that is called as the controlling community of the company. Compared with the controlling community, outside investors' controlling power over the company is weak and they have less information about the company. Controlling community can expropriate the funds of the company or the interests of minority shareholders. This behavior is usually called as expropriation, and the value that the controlling community extracts from the company and the minority shareholders is defined as agency costs. Therefore the benefits that outside investors can obtain from each share they hold tend to be less than the benefits that the controlling community can gain from each share they hold. In the following, we refer to the actual value that each share of the company brings for outside investors as the outside investor's value.

In order to analyze the mechanisms by which corporate governance affects the liquidity of corporate stocks, we construct a model framework that contains the heterogeneity of beliefs among different investors. In our theoretical model, investors in the market are divided into two types: informed traders and noise traders. The participation of noise traders in market transactions is an important condition for corporate governance to have the effect of improving the liquidity of the corporate stocks. In addition, our model also shows that the increased degree of noise traders' participation in market transactions will generally increase the liquidity of stocks of listed companies in the market, which is aligned with the conclusions of many classic literatures. In the assumptions of our model, informed traders can accurately grasp all the information about the company, while noise traders can accurately seize the information which is directly disclosed to the public or easily accessible. The scale of agency costs which are relatively secret and not directly disclosed can't be accurately known by them. Thus, informed traders can accurately know the value of the corporate equity, while noise traders may have biased estimations on its value. The noise traders in our model are essentially "semi-informed" traders rather than completely uninformed traders. Simply speaking, one important role of corporate governance in our model is to send noise traders a positive signal when they estimate the value of the corporate equity, thereby reducing the possibility that they overestimate the agency cost of the company and the extent to which their estimations deviate from the actual value that the equity can bring to them.

The core part of our model building can be divided into four steps. (1) Using discounted cash flow method, we deduce the estimations of different types of investors on the outside investors' value of the corporate equity based on their own information about the agency cost of the company. (2) Consider a situation where some holders need to sell a large amount of the corporate stocks in a short period of time. We assume that the transaction is conducted by open auction among all types of outside investors in the market, so an investor's estimation on the outside investors' value of the corporate equity determines the upper bound of his bids. We derive the probability distribution of the market price of the stock after the transaction is completed under market equilibrium. (3) Based on the probability distribution of its price, we design an index to measure the liquidity level of the corporate stocks, as establishes a relationship between its liquidity and corporate governance as well as the participation of noise traders in trading. (4) Based on these relationships, we analyze the effect of improving corporate governance on the liquidity of its stocks.

Finally, based on our theoretical model, we further demonstrates that corporate governance and the degree of noise traders' participation in trading have a synergistic effect on improving the liquidity of the corporate stocks : when the degree of noise traders' participation in trading increases, the liquidity of stocks of listed companies in the market will generally increase, and the liquidity of stocks of companies with better governance tends to get a greater increase than those with poor governance; additionally, compared with the periods when the degree of noise traders' participation in trading is low, improving the quality of corporate governance in the periods when this degree is high can improve the liquidity of the corporate stocks more greatly.

To make our analysis clear, we divide companies into two categories based on the power of the controlling community. For a company in the first category, there is no controlling community that have controlled the company for a long time, and they can't dictate the company absolutely. In this kind of company, the quality of corporate governance determines the scale of the company's agency costs. We call this type of company as general-type company. In the analysis of general-type companies, our model framework includes three roles of corporate governance. The first effect is that corporate governance can determine the maximum scale of the agency costs extracted by the controlling community. To be concrete, better governance can limit the scale of the company's agency costs to a relatively smaller range. The basis of containing this role in our model framework is the definition of corporate governance given by Shleifer and Vishny (1997). The second role which has been emphasized in La-Porta et al.(2002) is that better governance can increase the possibility that the controlling community's behavior of expropriation is captured and the intensity of the punishment for this behavior, which will accordingly affect their choice of the agency costs that they will extract from the company. The third role which has been mentioned above is that better corporate governance can send a positive signal to the noise traders who have incomplete information when they estimate the value of the corporate equity.

For a company in the second category, there is a controlling community that has firmly controlled the company for a long time, so the value of its agency costs tends to be fixed. In this kind of company, the controlling community have absolute control over the company and can even determine the arrangement of corporate governance in the company. Due to various reasons such as the requirements of the regulatory authorities, the controlling community also have to adjust the arrangement of corporate governance. However, they will try to ensure that the original scale of the agency costs that they can extract will not be restricted when they adjust the arrangement. Therefore, in such special companies corporate governance has a weaker impact on agency costs, while agency costs have a stronger influence on the quality of corporate governance. We call this kind of company as controlled-type company. Therefore, for controlled-type companies, the third aforementioned role of corporate governance should be included in our model framework, whereas the first and the second role should be excluded.

We arrange the structure of this paper as follows:

We first need to quantify the effect of corporate governance on its agency cost and an investor's estimation on the outside investors' value of the corporate equity based on his own information, as provide a theoretical basis for further research on the core issue of this paper. We completed this work in section 2. Part of the analysis draws on the theoretical framework of the existing literature.

Based on the conclusions of section 2, we construct our theoretical model in section 3 and 4.

In section 3, we established a theoretical model that contains the heterogeneity of investors' beliefs. By this model, we give an explanation for the mechanisms by which corpor ate governance generate an effect of improving the liquidity of the corporate stocks, as well as derive an important condition for apparent occurrence of this effect.

In Section 4, based on the model framework in Section 3, we further demonstrate that the strength of the effect that corporate governance improves the liquidity of the corporate stocks is affected by the degree of noise traders 'participation in stock trading. Likewise, the strength of the effect that noise traders' participation in trading improves the liquidity of stocks will also be affected by the quality of corporate governance. In other words, we prove that corporate governance and noise traders' participation have a synergistic effect on improving liquidity of stocks.

The analysis in parts 3 and 4 is targeted at general-type companies.

In Section 5, we focus on controlled-type companies and do relevant research. Our theoretical model indicates that the conclusions we draw in Sections 3 and 4 are still valid for this particular type of companies.

In Section 6, we summarize our analysis and give some policy recommendations.

## 2. The Impact of Corporate Governance on the Outside Investors' Value of the Corporate Equity

For general-type companies, the main process by which corporate governance affects the outside investors' value of their Equity is shown below:

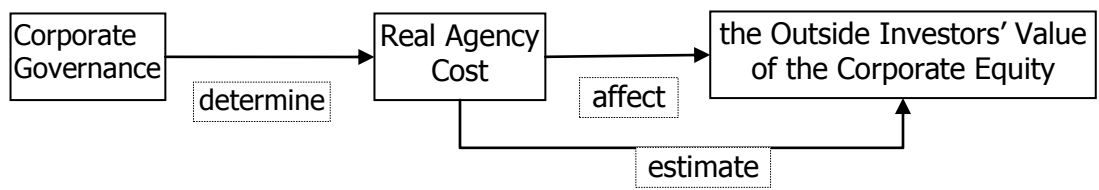

Figure 1

As shown in Figure 1, firstly, corporate governance determines the scale of the agency costs in a company; then, the agency cost will affect the outside investors' value of the corporate equity. Additionally, outside investors will also utilize their information about the agency cost of the company when estimating the outside investors' value. There are many important factors that can affect the outside investors' value of the corporate equity, and agency cost is one of them. The higher the agency cost of a company, that is, the controlling community's expropriation of the outside investors' interests is more serious, then the lower the outside investors' value of the corporate equity, that is, the earnings obtained by the outside investors will be reduced.

In section 2.1, we model the process of corporate governance determining actual agency costs, using an analysis framework similar to La Porta and Lopez-De-Silanes et al. (2002). In section 2.2, we model the process of agency costs affecting the outside investors' value of the corporate equity and an investors' estimation on the outside investors' value based on his own information. In this way, we quantitatively characterize the impact of corporate governance on the outside investors' value in this section and provide a basis for further research on the core issues of this paper.

2.1 Corporate Governance and Agency Costs

The analysis in this paper proceeds under continuous time settings. Consider a representative company in the economy. Referring to the assumptions given by Morellec and Nikolov et al. (2012, 2018), We set the production function of the company at the time of t as:

$y_t = A_t k_t^\alpha, \quad \alpha < 1.$

$y_t$ is the total output of the company at the time of t, and $k_t$ is the total amount of funds which it invests in production. $\delta$ denotes the company's capital depreciation rate, and r denotes the borrowing rate. δ and r are all constants. $A_t$ represents the technical shock of the company at the time of t, and it is a random variable. Like the assumption of Morellec and Nikolov et al. (2012, 2018), the products produced by the company at the time of t will sell out simultaneously and one unit product is sold at a unit price. The cost of producing one unit product is $C_t$ which can be treated as the cost shock at the time of t and is also a random variable. We define $Z_t \triangleq A_t - C_t$. Similar to the assumption of Morellec and Nikolov et al. (2012, 2018), we assume that the stochastic process $\{Z_t\}_{t \geq 0}$ is governed by a stochastic differential equation which is:

$$dZ_t = \mu_z Z_t dt + \sigma_z Z_t dW_t \qquad (2.1)$$

$\{W_t\}_{t \geq 0}$ is a standard Brownian motion which reflects the uncertainty of productivity and cost shocks. Therefore the profit of the company at the time of t is:

$$Pft_t = Z_t k_t^\alpha - (r + \delta) k_t$$

Now, we introduce the agent cost which is an important variable into our model in the form that has been used by morelec, Nikolov, et al. (2018), La porta, Lopez de silanes, et al. (2002), Shleifer, wolfenzon (2002). They assume that the controlling community can extract a constant share ρ of the company's profits at each moment and the share ρ is regarded as the agency cost, where $\rho \in [0,1)$. To be concrete, it means that the cash flow which the controlling community expropriates from the company at any time t is $\rho \cdot Pft_t$ and the remaining profit $(1-\rho) \cdot Pft_t$ is then distributed among all shareholders of the company on a pro rata basis. We suppose the controlling community holds a fraction θ of the equity of the company totally. Here ϴ is exogenous and determined by its history or its life-cycle. Thus, the total value of the cash that the controlling community can obtain at any time t is $\rho \cdot Pft_t + \theta \cdot (1-\rho) \cdot Pft_t$.

Next, we introduce the factor of corporate governance into the model. In order to reflect the inhibitory effect of corporate governance on agency costs, we assume that $\rho \in [0, c_m]$, that is, the extraction proportion of the controlling community can't exceed $c_m$, where $c_m \epsilon [0,1)$. Therefore, $c_m$ is a parameter which reflects the quality of corporate governance. For a company, the smaller $c_m$ is, the better its corporate governance is and the stronger the ability to restrain its agency costs. The literature we mentioned above assumes that ρ and $c_m$ are constant and don't change with time. This assumption is used in this paper for the convenience of analysis. In order to fully reflect the role of corporate governance in our model, we use the methods of La porta, Lopez de silanes, et al. (2002) to consider the punishment given to the controlling community for their illegally extracting corporate funds. Suppose that in a company, the share of profits extracted by the controlling community is ρ,

The probability of their expropriation being caught and revealed is $p(c_m) \cdot \rho$, where $p'(c_m) < 0$. That is, the larger the expropriation ratio ρ that the controlling community chooses, the greater the possibility that they will be punished; the worse the company's corporate governance, the greater the possibility that they will be punished. The punishments here include not only the fines imposed on them and the loss of their positions in the company, but also the damage to their reputation in the industry.

After their expropriation is caught and revealed, they were punished with $d(c_m)$ for each unit of the value they extracted, where $d'(c_m) < 0$. That is, the worse the company's corporate governance, the less the punishment imposed on them.

Therefore, the expectation of the value that the controlling community is fined at time t is: $\left(\rho^2 / f(c_m)\right) \cdot Pft_t$, where $f(c_m) = \frac{1}{p(c_m) \cdot d(c_m)}$. Obviously, $f'(c_m) > 0$. The expectation of the value obtained by the controlling community at this time is:

$$\rho \cdot Pft_t + \theta \cdot (1-\rho) \cdot Pft_t - \frac{\rho^2}{f(c_m)} \cdot Pft_t$$

$$= \left[\rho + (1-\rho)\theta - \frac{\rho^2}{f(c_m)}\right] \cdot (Z_t k_t^\alpha - (r+\delta)k_t)$$

Because the controlling community has control over the company, they can determine the total amount of capital $k_t$ that the company puts into production at each time and then the output of the company. Their goal is to maximize their benefits at every time. Therefore, at any time t, they should choose the optimal $\rho$ and $k_t$ to solve the optimization problem:

$$Q_C(t) = \max_{\substack{0 \leq \rho \leq c_m \\ k_t \geq 0}} \left[\rho + (1-\rho)\theta - \frac{\rho^2}{f(c_m)}\right] \cdot [Z_t k_t^\alpha - (r+\delta)k_t] \qquad (2.2)$$

In fact, solving this optimization problem is equivalent to solving the following two optimization problems:

$$\max_{0 \leq \rho \leq c_m} \rho + (1-\rho)\theta - \frac{\rho^2}{f(c_m)} \qquad (2.3)$$

and

$$\max_{k_t \geq 0} Z_t k_t^\alpha - (r+\delta)k_t. \qquad (2.4)$$

The first order condition of the optimization problem (2.3) is:

$$1 - \theta - \frac{2\rho}{f(c_m)} = 0$$

Then $\rho$ becomes a function of the parameter $c_m$ of corporate governance:

$$\rho(c_m) = \frac{1-\theta}{2} \cdot f(c_m) \qquad (2.5)$$

Then $\rho$ is an increasing function of the parameter $c_m$, that is, the company with poor governance has relatively high agency cost. This is intuitive and aligned with the basic theory of corporate governance. By equation (2.5), we can characterize the impact of corporate governance on the company's agency cost.

The solution to the optimization problem (2.4) is:

$$k_t^* = \left(\frac{\alpha}{r+\delta}\right)^{\frac{1}{1-\alpha}} \cdot Z_t^{\frac{1}{1-\alpha}} \qquad (2.6)$$

Solution (2.6) maximizes the expectation of the benefits of the controlling community at the time of t, as well as the profits of the company at this time.

2.2 Agency Cost and Outside Investor's Value of Stocks

When the controlling community adopts the optimal choice (2.5) and (2.6), the expectation of their benefits is:

$$Q_C(t) = \left[\rho + (1-\rho)\theta - \frac{\rho^2}{f(c_m)}\right] \cdot (Z_t(k_t^*)^\alpha - (r+\delta)k_t^*)$$

If the number of the company's shares is $\bar{S}$, then the benefit for each share held by the controlling community at the time of t is:

$$\frac{Q_C(t)}{\theta \cdot \bar{S}} = \left[1 - \rho + \frac{\rho \cdot f(c_m) - \rho^2}{\theta \cdot f(c_m)}\right] \frac{(1-\alpha)}{\bar{S}} \left(\frac{\alpha}{r+\delta}\right)^{\frac{\alpha}{1-\alpha}} \cdot Z_t^{\frac{1}{1-\alpha}}$$

The total income of all the outside investors at the time of t is:

$$Q_U(t) = [(1-\rho) \cdot (1-\theta)] \cdot (Z_t(k_t^*)^\alpha - (r+\delta)k_t^*)$$
$$= (1-\rho)(1-\theta)(1-\alpha)\left(\frac{\alpha}{r+\delta}\right)^{\frac{\alpha}{1-\alpha}} \cdot Z_t^{\frac{1}{1-\alpha}}$$

Then the benefit for each share held by an outside investor at the time of t is:

$$\frac{Q_U(t)}{(1-\theta)\bar{S}} = (1-\rho)\frac{(1-\alpha)}{\bar{S}}\left(\frac{\alpha}{r+\delta}\right)^{\frac{\alpha}{1-\alpha}} \cdot Z_t^{\frac{1}{1-\alpha}}$$

Due to the existence of agency costs, the amount that the controlling group would benefit more than outside investors from each share held is:

$$\Delta Q(t) = \frac{Q_C(t)}{\theta \cdot \bar{S}} - \frac{Q_U(t)}{(1-\theta)\bar{S}}$$
$$= \left[\frac{\rho \cdot f(c_m) - \rho^2}{\theta}\right] \frac{(1-\alpha)}{\bar{S}}\left(\frac{\alpha}{r+\delta}\right)^{\frac{\alpha}{1-\alpha}} \cdot Z_t^{\frac{1}{1-\alpha}} > 0$$

$$= \left[\frac{(1-\theta^2) \cdot f^2(c_m)}{4\theta}\right] \frac{(1-\alpha)}{\bar{S}} \left(\frac{\alpha}{r+\delta}\right)^{\frac{\alpha}{1-\alpha}} \cdot Z_t^{\frac{1}{1-\alpha}} > 0$$

$\Delta Q(t)$ can be regarded as the benefit of control which has been extensively studied by existing literature. It can be easily deduced that $\Delta Q(t)$ decreases with the improvement of corporate governance (i.e., the decrease of $c_m$).

For the convenience of later discussion, we define a stochastic process:

$$X_t \triangleq \frac{(1-\alpha)}{\bar{S}} \left(\frac{\alpha}{r+\delta}\right)^{\frac{\alpha}{1-\alpha}} Z_t^{\frac{1}{1-\alpha}}$$

So at the time of t, the benefit that each share brings to an outside investor is:

$$EP_t \triangleq \frac{Q_U(t)}{(1-\theta)\bar{S}} = (1-\rho(c_m)) \cdot X_t \qquad (2.7)$$

According to the expression (2.1) of the stochastic process $\{Z_t\}_{t\geq 0}$, using the Ito formula can derive that the stochastic process $\{X_t\}_{t\geq 0}$ is governed by the following stochastic differential equation:

$$dX_t = \mu X_t dt + \sigma X_t dW_t, \quad X_0 = \frac{(1-\alpha)}{\bar{S}} \left(\frac{\alpha}{r+\delta}\right)^{\frac{\alpha}{1-\alpha}} \cdot Z_0^{\frac{1}{1-\alpha}}$$

where $\mu = \frac{\mu_z}{1-\alpha} + \frac{\sigma_z^2}{2} \cdot \frac{\alpha}{(1-\alpha)^2}$ and $\sigma = \frac{\sigma_z}{1-\alpha}$. $Z_0$ and $X_0$ represent the initial states of the stochastic processes $\{Z_t\}_{t\geq 0}$ and $\{X_t\}_{t\geq 0}$, respectively.

We derive the solution of the above stochastic differential equation:

$$X_t = X_0 \, e^{(\mu-\frac{\sigma^2}{2})t} \, e^{\sigma W_t}$$

It's generally recognized that investors always estimate the value of the corporate equity based on the relevant information they have obtained. We assume that all outside investors can accurately know the company's parameters which are often directly disclosed to the public or easily accessible, including equity concentration $\theta$, $\mu_z$, $\sigma_z$, $\mu$, $\sigma$, $\alpha$, $\delta$, $\bar{S}$ and $r$. Additionally, most of the information about the governance arrangement of listed companies is publicly disclosed, and outside investors can learn about the quality of corporate governance of these companies through the information, so we assume that all of them can accurately know the parameter $c_m$ of corporate governance.

On the other hand, we assume that some of them are not able to precisely know the scale of the agency cost $\rho$ because it's relatively secret and not directly disclosed. Thus there may exist heterogeneity between different investors' estimations on the agency cost of one company. In our model, investors who can accurately know the scale of the company's agency cost are regarded as informed traders, and the other investors are regarded as noise traders. That is, the difference between noise traders and informed traders in our model lies in the accuracy of knowing the scale of agency costs.

We set a representative outside investor's estimation on the agency cost of the company as $\hat{\rho}$ which may be biased from the real value $\rho$. Then by equation (2.7) we deduce that he tends to estimate the income brought by each share held by himself at the time of t as $\widehat{EP_t} = (1-\hat{\rho}) \cdot X_t$. Now we calculate $V(t, \hat{\rho})$ which is defined as his estimation on the value of each share of the company at the time of t. We calculate $V(t, \hat{\rho})$ by discounting all the cash flows obtained by holding one share after the time of t, and the discount rate of outside investors is set to $\gamma$.

$$V(t, \hat{\rho}) = E_t\left[\int_t^{+\infty} e^{-\gamma(s-t)} \widehat{EP_t} \, ds\right]$$

$$= E_t\left[\int_t^{+\infty} e^{-\gamma(s-t)} (1-\hat{\rho}) \cdot X_s \, ds\right]$$

$$= \int_t^{+\infty} E_t\left[e^{-\gamma(s-t)} (1-\hat{\rho}) \cdot X_s\right] ds$$

$$= \int_t^{+\infty} e^{-\gamma(s-t)} (1-\hat{\rho}) E_t\left[X_0 \, e^{(\mu-\frac{\sigma^2}{2})s} \, e^{\sigma W_s}\right] ds$$

$$= \frac{(1-\hat{\rho})}{\gamma - \mu} X_0 \, e^{(\mu - \frac{\sigma^2}{2})t} \, e^{\sigma W_t} \tag{2.8}$$

The second equal sign of the above formula is derived by Fubini's theorem. From the perspective of discounted cash flow method, the actual outside investor's value of each share at this time should be:

$$V(t, \rho) = \frac{(1-\rho)}{\gamma - \mu} X_0 \, e^{(\mu - \frac{\sigma^2}{2})t} \, e^{\sigma W_t} \tag{2.9}$$

since the real agency cost of the company is ρ. All the words "value" that occur in the rest of this paper refer to the outside investor's value of the corporate stocks.

Remark: The agency cost itself is a piece of information that is difficult to acquire directly. Although we follow La porta, Lopez de silanes, et al. (2002) to express the real agency cost as a function of the parameter $c_m$ of corporate governance by equation (2.5), in fact, noise traders usually can't use $c_m$ to precisely infer the agency cost of the company. There are two main reasons. (1) In reality, the determinants of a company's agency cost are complex and diverse, including the history of the company and its life-cycle. The ways in which various factors affect the agency cost are also changeable. Formula (2.5) is only a simplified and imprecise description of the scale of agency cost. (2) Due to poor information and knowledge, it is difficult for noise traders to fully seize the factors that can influence the agency cost and the way in which these factors jointly determine the agency cost.

## 3. the Effect of Corporate Governance on the Liquidity of Stocks

3.1 Investors' Heterogeneous Beliefs and the Basic Setting of Our Model

In this section, taking the existence of noise traders in market transactions, we analyze the mechanisms by which corporate governance affect the liquidity of the corporate stocks. The analysis in this section is for general-type companies, and we will give the corresponding conclusions for controlled-type companies in section 5.

Now We consider a representative listed company with the parameters described in Section 2. We assume that there exist both informed traders and noise traders in the market. The number of informed traders and noise traders in the market is $N_I$ and $N_0$, respectively. $N_I$ and $N_0$ are both constants that do not change with time. Informed traders have complete information, including the real agency cost of the company and the actual outside investor's value of the corporate equity. Because they are completely informed, it's difficult for them to lose money in trading and they tend to make some profits. Therefore, we assume that all of them are always willing to use their information to participate in the trading of the corporate stocks. Noise traders without complete information bear a certain degree of risk of loss in trading, so not all noise traders are always willing to participate in trading. Furthermore, the number of people willing to participate in trading is random and potentially changes over time. Thus, following the assumptions in Easley, Kiefer and O'hara et al. (1996), we suppose that the number of noise traders willing to participate in market transactions over time follows a Poisson process with arrival rate λ. The larger the arrival rate λ, the higher the degree of noise traders' participation in trading and the more active they are in the market. In the following we refer to λ as the degree of noise traders' participation in trading.

We focus on a generalized situation in which a shareholder of the company has to sell out the company's stocks held by him at the time of $T_2$ because of his urgent need for cash. The total number of the shares held by him is $n \cdot M$. For the convenience of analysis, we assume that each investor in the market can only buy n shares of the company until the time of $T_2$. Therefore, at the time of $T_2$, the shareholder will sell his shares by M deals that he will make with different investors, and each deal consists of n shares of the company. The price of each deal is determined by the open auction between investors who a re willing to participate in trading at that time. We assume that at a certain time point $T_1$ before the time point of $T_2$, some transactions on the company's stocks have occurred in the market, as have caused that all the noise traders who are willing to participate at the time of $T_1$ have purchased ample shares in these transactions and left the market at the time of $T_1$. Between the time of $T_1$ and $T_2$, no transaction on the stocks of the company occurs. So the investors who will participate in the open auction at the time of $T_2$ include:
(1) all informed traders in the market, the number of which is $N_I$;
(2) noise traders willing to participate in trading in the time interval $[T_1, T_2]$, the number of which is denote by $N_U$.

According to our assumption, $N_U$ obeys a Poisson distribution with parameter $\lambda \cdot \Delta T$, where $\Delta T = T_2 - T_1$. $\Delta T$ is the length of the time interval between these two transactions. We define $N \triangleq N_I + N_U$. In addition, we assume that no trader in the market can become a large shareholder of the company and join the controlling community of the company through these transactions, so they only care about the outside investor's value of each share at the time of $T_2$. Let $\overline{V}$ denote the value of each share when there is no agency cost in the company. By equation (2.9), we derive

$$\overline{V} = V(T_2, 0) = \frac{1}{\gamma - \mu} X_0 \, e^{\left(\mu - \frac{\sigma^2}{2}\right)T_2} \, e^{\sigma W_{T_2}}$$

If the controlling community extract the cash flow of the company to the maximum extent permitted by the constraints of governance mechanisms of the company, that is, their expropriation ratio reaches the upper bound $c_m$, by equation (2.9) the value of each share should be:

$$\underline{V} \triangleq V(T_2, c_m) = \frac{(1 - c_m)}{\gamma - \mu} X_0 \, e^{\left(\mu - \frac{\sigma^2}{2}\right)T_2} \, e^{\sigma W_{T_2}} = (1 - c_m) \cdot \overline{V}$$

By equation (2.9) again, the real value of each share should be:

$$V(\rho) \triangleq V(T_2, \rho) = \frac{(1 - \rho)}{\gamma - \mu} X_0 \, e^{\left(\mu - \frac{\sigma^2}{2}\right)T_2} \, e^{\sigma W_{T_2}} = (1 - \rho) \cdot \overline{V}$$

$V(\rho)$ can also be seen as the fair value of each share. Informed traders deem the value of each share to be $V(\rho)$ correctly. If a noise trader estimates the agency cost $\rho$ as $\hat{\rho}$, by formula (2.8) we deduce that his estimation on the value of each share at the time of $T_2$ should be:

$$\widehat{V} \triangleq V(T_2, \hat{\rho}) = \frac{(1 - \hat{\rho})}{\gamma - \mu} X_0 \, e^{\left(\mu - \frac{\sigma^2}{2}\right)T_2} \, e^{\sigma W_{T_2}} = (1 - \hat{\rho}) \cdot \overline{V}$$

It can be inferred $\hat{\rho} \in [0, c_m]$, since noise traders precisely know the parameter $c_m$ of corporate governance. Because their estimations tend to be biased and the deviations between their estimations and the real size $\rho$ are random, we assume that all noise traders' estimations on the size of agency cost obeys a uniform distributed on the interval of $[0, c_m]$ and are mutually independent. Therefore, $\widehat{V}$ obeys a uniform distributed on the interval of $[\underline{V}, \overline{V}]$ and are mutually independent.

3.2 Equilibrium Price of Each Share

In the process of this open auction which will complete at the time of $T_2$, the following conditions usually hold.

**Condition 1**. Every investor can observe the bids of all the traders in the market.
**Condition 2**. Before the time of $T_2$, the number of times each trader can modify his bid is unlimited.
**Condition 3**. Noise traders participating in the open auction don't know whether there are informed traders in the market.
**Condition 4**. The seller doesn't know any bidder's estimation on the value of each share, i.e. $\widehat{V}$.
**Condition 5**. At the time of $T_2$, the M deals are concluded one by one in the form of open auction. Traders don't ex ante know how many shares the seller will sell.

**Condition 6**. Use P to denote the price to pay for one share. If an investor believes that buying a stock makes him neither lose nor gain based on his information (i.e. $\widehat{V} = P$), he won't choose to buy it.

We also need to consider a special case in which the number of the noise traders who are willing to participate in trading is so small or the number of the shares that the shareholders need to sell is so large that $N_U + N_I < M$. This case essentially means that selling pressure in the market is heavy and the liquidity that noise traders can provide is not enough. In this case, all traders willing to participate in trading have bought n shares and quit the market, whereas the seller still holds some shares that haven't been sold. We make the following assumptions.

**Assumption 1**. If the seller who has to sell out his stocks at the time of $T_2$ quotes an asking price of $\underline{V}$, this price can effectively attract other noisy traders who were not willing to participate in trading to buy the remaining shares from the seller at this time.

Because in the belief of noise traders in the market, the probability that the actual value of each share is lower than $\underline{V}$ is 0, buying at the price of $\underline{V}$ is unlikely to make them lose and may bring them earnings. Thus, Assumption 1 is reasonable.

We use the following proposition to explain what the market price $P_{T2}$ of each share will be after all the M deals are concluded under market equilibrium at the time of $T_2$. In fact, $P_{T2}$ is the equilibrium price of each share in the last deal (i.e. the M-th deal).

**Proposition 1**. Suppose that there are totally N traders bidding in this open auction, where N is a constant positive integer. Their estimations on the outside investor's value of each share of the company in descending order is: $\widehat{V}^{(1)} > \widehat{V}^{(2)} > \cdots > \widehat{V}^{(N)}$. Suppose that the M deals are concluded under market equilibrium, then

(1) $P_{T2} = \widehat{V}^{(M+1)}$ if $N > M$;
(2) $P_{T2} = \underline{V}$ if $N \leq M$.

*Proof.*

For the convenience of our discussion, the trader who estimates the value of each share as $\widehat{V}^{(i)}$ is recorded as the trader $I_{(i)}$ for $i = 1, 2, \ldots, N$.

In the state of market equilibrium, any trader who changes his bid alone won't increase his benefits. In other words, the market is not in equilibrium if a trader in the market can increase his earnings by modifying his own bid. Now we demonstrate that the first deal will be closed by the trader $I_{(1)}$ at the price of $\widehat{V}^{(2)}$ for each share.

The first deal won't be concluded at a price of $\widetilde{V}_1$ which is greater than $\widehat{V}^{(2)}$, otherwise the market can't be in equilibrium. (1) The price of $\widetilde{V}_1$ for each share is unlikely to be bidden by a trader $I_{(j)}$ where $j > 1$, as means that traders other than the trader $I_{(1)}$ won't bid $\widetilde{V}_1$ for each share. The trader $I_{(j)}$ believes that buying each share at the price of $\widetilde{V}_1$ would bring him a loss, because his estimation on the value of each share satisfies that $\widehat{V}^{(j)} \leq \widehat{V}^{(2)} < \widetilde{V}_1$. (2) It is also impossible that the price of $\widetilde{V}_1$ for each share is bidden by the trader $I_{(1)}$, because there still exists some space for him to lower his bid for the purpose of increasing his benefits. For instance, it's profitable for him to lower his bid to a certain price of $\widetilde{\widetilde{V}}_1$ which satisfies that $\widehat{V}^{(2)} \leq \widetilde{\widetilde{V}}_1 < \widetilde{V}_1$. By the bid of $\widetilde{\widetilde{V}}_1$ for each share, he can also close the first deal because the other traders whose estimations on the value of each share are not larger than $\widehat{V}^{(2)}$ won't give a bid larger than $\widetilde{\widetilde{V}}_1$.

Additionally, the first deal won't be concluded at a price of $\dot{V}_1$ which is lower than $\widehat{V}^{(2)}$ for each share, otherwise the market can't be in equilibrium. There are at least two traders whose estimations on the value of each share is greater than $\dot{V}_1$, since $\widehat{V}^{(1)} > \widehat{V}^{(2)} > \dot{V}_1$. (1) If the trader who bids $\dot{V}_1$ for each share and will conclude the deal is the trader $I_{(2)}$, the trader $I_{(1)}$ can give a higher bid of $\ddot{V}_1$ which satisfies that $\widehat{V}^{(1)} > \ddot{V}_1 > \dot{V}_1$. By this bid, the trader $I_{(1)}$ can replace the trader $I_{(2)}$ to close the deal and get a profit of $M \cdot (\widehat{V}^{(1)} - \ddot{V}_1)$ in his belief. (2) If the trader who bids $\dot{V}_1$ for each share and will conclude the deal isn't the trader $I_{(2)}$, the trader $I_{(2)}$ can give a higher bid of $\ddot{V}_1$ which satisfies that

$\widehat{V}^{(2)} > \ddot{V}_1 > \dot{V}_1$. By this bid, the trader $I_{(2)}$ can replace the trader who bids $\dot{V}_1$ for each share to close the deal and get a profit of $M \cdot (\widehat{V}^{(2)} - \ddot{V}_1)$ in his belief.

Therefore, the first deal can only be concluded at the price of $\widehat{V}^{(2)}$ for each share and the trader who gives this bid and closes this deal must be the trader $I_{(1)}$. It is not possible for the other traders to give bids which are greater than or equal to $\widehat{V}^{(2)}$ for each share. If the trader $I_{(1)}$ lowers his bid to $V'$ (i.e. $V' < \widehat{V}^{(2)}$) for each share, the trader $I_{(2)}$ will give a bid which is higher than $V'$ but lower than $\widehat{V}^{(2)}$ for each share to replace him to buy the stocks and make a profit. If the trader $I_{(1)}$ raises his bid, his profit will decrease. After the first deal is concluded, the trader $I_{(1)}$ will quit the market and not participate in the transactions.

According to a similar analysis, we can deduce that the second deal will be concluded by the trader $I_{(2)}$ at the price of $\widehat{V}^{(3)}$ for each share. It can be proved by mathematical induction that for any integer k that satisfies $k \leq min\{N-1, M\}$, the k-th deal will be closed by the trader $I_{(k)}$ at the price of $\widehat{V}^{(k+1)}$ for each share. Therefore, if $N > M$, we can derive that $P_{T2} = \widehat{V}^{(M+1)}$ because the last deal (i.e. the M-th deal) will be concluded at the price of $\widehat{V}^{(M+1)}$ for each share.

If N = M, the k-th deal will be concluded by the trader $I_{(k)}$ at the price of $\widehat{V}^{(k+1)}$ for k = 1,2, ..., M-1. Then all of them will quit the market, which results that only one trader who is the trader $I_{(M)}$ will bid for the last n shares. His optimal strategy is to give a bid of $\underline{V}$ for each share, which ensures that he can conclude the last deal. (1) If he gives a bid that is lower than $\underline{V}$, the seller can then quote an asking price of $\underline{V}$ and sell the last n shares to another trader in the market according to Assumption 1. (2) On the other hand, giving a bid which is higher than $\underline{V}$ can reduce his profit. Thus, we deduce that $P_{T2} = \underline{V}$ if N = M.

If $N < M$, it's obvious that $P_{T2} = \underline{V}$ by Assumption 1.

□

By our settings, $P_{T2}$ is obviously a random variable, as is also consistent with reality.

If the selling pressure at the time of $T_2$ is so small that $M < N_I$, it's obvious that $P_{T2} \geq V(\rho)$ by Proposition 1. This means that the seller can sell each share held by him at a price that is not lower than the fair value of each share. Since this paper focus on the liquidity of stocks, we assume that the selling pressure at this time is so large that $M \geq N_I$ in the rest of this paper.

3.3 Our Measurement of Liquidity of Stocks

It has been defined above that $V(\rho)$ is the actual outside investor's value of each share of the company at the time of $T_2$ and $P_{T2}$ is the market price of each share after the shareholder sells out all the $M \cdot n$ shares held by him at this time. If $P_{T2} \leq V(\rho)$, we define

$$s \triangleq 1 - \frac{P_{T2}}{V(\rho)}$$

In fact, $s$ which is obviously a random variable denotes the discount rate of the market price of each share at the time of $T_2$ relative to its real value. In the following, we refer to $s$ as the discount rate. We define $\bar{s} \triangleq \frac{c_m - \rho}{1 - \rho}$. Because it's obvious that $P_{T2} \geq \underline{V}$ according to Proposition 1, we derive

$$s \leq 1 - \frac{\underline{V}}{V(\rho)} = 1 - \frac{1 - c_m}{1 - \rho} = \bar{s}$$

So the range of $s$ is $[0, \bar{s}]$. Keeping the time interval $\Delta T$, the degree of noise trader's participation $\lambda$ and the trading volume $n \cdot M$ fixed, the larger the discount rate $s$, the larger the extent to which the price of each share deviates downwards from its fair value after the trading. A discount rate of $s$ suggests that the seller obtained only 1-s of the real value of each share in the last deal, that is, he lost $s \cdot V(\rho)$ for each share. In order to characterize the liquidity level of stocks in our model framework, for any constant number $s_0$ in $[0, 1)$, we can define

$$F_{s_0}(c_m, \lambda) \triangleq \Pr\left[s_0 < 1 - \frac{P_{T2}}{V(\rho)}\right] = \Pr[P_{T2} < (1-s_0)V(\rho)]$$

and

$$ILL_{s_0}(c_m, \lambda) \triangleq \begin{cases} -\ln F_{s_0}(c_m, \lambda), & \text{if } F_{s_0}(c_m, \lambda) > 0 \\ +\infty, & \text{if } F_{s_0}(c_m, \lambda) = 0 \end{cases}$$

$-ILL_{s_0}(c_m, \lambda)$ which can be seen as a function of $s_0$, $c_m$ and $\lambda$ is the natural logarithm of the probability that the discount rate of the market price of each share relative to its fair value is larger than $s_0$ after the trading. For a fixed $s_0$, the greater the $ILL_{s_0}(c_m, \lambda)$, the higher the level of liquidity of the stocks.

Consider two specific listed companies which are company A and company B. For any fixed $s_0$ in $[0,1)$, their levels of liquidity are $ILL_{s_0}(c_m^A, \lambda)$ and $ILL_{s_0}(c_m^B, \lambda)$ respectively. If $ILL_{s_0}(c_m^A, \lambda) > ILL_{s_0}(c_m^B, \lambda)$ holds for any $s_0 \in [0,1)$, one can claim that the stocks of company A are more liquid than that of company B.

For a representative trader whose estimation on the value of each share at the time of $T_2$ is $\widehat{V}$, use $g_{s_0}(c_m)$ to denote the probability that the discount rate of his estimation $\widehat{V}$ relative to the real value $V(\rho)$ is larger than $s_0$, i.e.

$$g_{s_0}(c_m) \triangleq \Pr[\widehat{V} < (1-s_0)V(\rho)],$$

where $s_0$ can be any constant number in $[0,1)$. For any $s_0 \in [0, \bar{s})$, we derive

$$g_{s_0}(c_m) = \frac{(1-s_0)V(\rho) - V(c_m)}{\overline{V} - V(c_m)} = 1 - \frac{\rho + s_0 - s_0\rho}{c_m}. \tag{3.1}$$

For any $s_0 \in [\bar{s}, 1)$, it's obvious that $g_{s_0}(c_m) = 0$.

Now We calculate our liquidity indicator of the stocks under the case of the open auction that we have elaborated in Section 3.1. If $M < N_I$, then $M < N_I + N_U = N$. All informed traders estimate the value of each share as $V(\rho)$. Thus, by Proposition 1 we can deduce that $P_{T2} = \widehat{V}^{(M+1)} \geq V(\rho)$, which means that $\Pr\{1 - \frac{P_{T2}}{V(\rho)} \leq 0\} = 1$. So, For any $s_0 \in [0,1)$, we derive

$$F_{s_0}(c_m, \lambda) = 0.$$

This means that trading in small quantities doesn't tend to cause a sharp decline in the market price of the stocks. If $M \geq N_I$, we have the following conclusion:

**Proposition 2.** At the time of $T_2$, for any $s_0 \in [0, \bar{s})$, the probability that the discount rate of the market price of each share relative to its fair value is larger than $s_0$ is:

$$F_{s_0}(c_m, \lambda) = \sum_{i=0}^{m} \frac{(\lambda \Delta T)^i}{i!} e^{-(\lambda \Delta T)}$$
$$+ \sum_{i=m+1}^{+\infty} \sum_{j=0}^{m} \left\{ \frac{(\lambda \Delta T)^i}{j!(i-j)!} e^{-(\lambda \Delta T)} [1 - g_{s_0}(c_m)]^j [g_{s_0}(c_m)]^{i-j} \right\}$$

where $m \triangleq M - N_I$. For any $s_0 \in [\bar{s}, 1)$,

$$F_{s_0}(c_m, \lambda) = 0.$$

*Proof.*
(1) Focus on the case in which $s_0 \in [0, \bar{s})$. If $N_U$ which is the number of noise traders participating in bidding satisfies that $N_U \leq m$, then $N$ which is the total number of traders participating it satisfies that $N \leq M$ and we can derive $P_{T2} = \underline{V} < (1-s_0)V(\rho)$ by Proposition 1.

If $N_U > m$, then $N > M$ and we can derive $P_{T2} = \widehat{V}^{(M+1)}$ by Proposition 1. We still set that the estimations on each share of all the $N$ traders in descending order is:

$$\widehat{V}^{(1)} > \widehat{V}^{(2)} > \cdots > \widehat{V}^{(N)}.$$

We additionally set that the estimations on each share of all the $N_U$ noise traders among all the $N$ traders in descending order is:

$$\hat{v}^{(1)} \geq \hat{v}^{(2)} \geq \cdots \geq \hat{v}^{(N_U)}.$$

Notice that all the informed traders estimate the value of each share as its actual value $V(\rho)$ which satisfies that $V(\rho) \geq (1-s_0)V(\rho)$. So, if $\widehat{V}^{(M+1)} < (1-s_0)V(\rho)$, we can derive that $\hat{v}^{(m+1)} = \widehat{V}^{(M+1)}$ and it follows that $\hat{v}^{(m+1)} < (1-s_0)V(\rho)$. Likewise, if $\hat{v}^{(m+1)} < (1-s_0)V(\rho)$, it's obvious that $\widehat{V}^{(M+1)} = \hat{v}^{(m+1)}$ and it follows that $\widehat{V}^{(M+1)} < (1-s_0)V(\rho)$. Therefore, we can deduce that $\{\widehat{V}^{(M+1)} < (1-s_0)V(\rho)\} \Leftrightarrow \{\hat{v}^{(m+1)} < (1-s_0)V(\rho)\}$.

For convenience of our deduction, we define that $\hat{v}^{(0)} \triangleq V(\rho)$.

$\Pr[P_{T_2} < (1-s_0)V(\rho)]$

$= \Pr[N_U \le m] + \sum_{i=m+1}^{+\infty} \Pr[\{N_U = i\} \cup \{\widehat{V}^{(M+1)} < (1-s_0)V(\rho)\}]$

$= \sum_{i=0}^{m} \Pr[N_U = i] + \sum_{i=m+1}^{+\infty} \Pr[N_U = i] \cdot \Pr[\widehat{V}^{(M+1)} < (1-s_0)V(\rho) \mid N_U = i]$

$= \sum_{i=0}^{m} \Pr[N_U = i] + \sum_{i=m+1}^{+\infty} \Pr[N_U = i] \cdot \Pr[\hat{v}^{(m+1)} < (1-s_0)V(\rho) \mid N_U = i]$

$= \sum_{i=0}^{m} \Pr[N_U = i]$

$+ \sum_{i=m+1}^{+\infty} \left\{ \Pr[N_U = i] \cdot \sum_{j=0}^{m} \Pr[\hat{v}^{(j)} \ge (1-s_0)V(\rho) > \hat{v}^{(j+1)} \mid N_U = i] \right\}$

$= \sum_{i=0}^{m} \frac{(\lambda \Delta T)^i}{i!} e^{-(\lambda \Delta T)}$

$+ \sum_{i=m+1}^{+\infty} \left\{ \frac{(\lambda \Delta T)^i}{i!} e^{-(\lambda \Delta T)} \cdot \sum_{j=0}^{m} \binom{i}{j} [1 - g_{s_0}(c_m)]^j [g_{s_0}(c_m)]^{i-j} \right\}$

(2) By proposition 1, it's obvious that $P_{T_2} \ge \underline{V}$. Thus, for any constant $s_0$ in $[\bar{s}, 1)$, we derive

$$F_{s_0}(c_m, \lambda) = \Pr\left[s_0 < 1 - \frac{P_{T_2}}{V(\rho)}\right] \le \Pr\left[s_0 < 1 - \frac{\underline{V}}{V(\rho)}\right] = \Pr[s_0 < \bar{s}] = 0$$

$\square$

For convenience of the following discussion, we define that

$$K(g, L) \triangleq \sum_{i=0}^{m} \frac{L^i}{i!} e^{-L} + \sum_{i=m+1}^{+\infty} \sum_{j=0}^{m} \left[ \frac{L^i e^{-L}}{j! (i-j)!} (1-g)^j g^{i-j} \right]. \qquad (3.2)$$

Then, for any constant $s_0$ in $[0, \bar{s})$, we derive

$$F_{s_0}(c_m, \lambda) = \begin{cases} K(g_{s_0}(c_m), \lambda \Delta T), & \text{if } s_0 \in [0, \bar{s}) \\ 0, & \text{if } s_0 \in [\bar{s}, 1) \end{cases}$$

It can be seen from Proposition 2 that $F_{s_0}(c_m, \lambda)$ and $ILL_{s_0}(c_m, \lambda)$ have nothing to do with the starting and ending time points $T_1$ and $T_2$, but are related to the interval time $\Delta T$ between them.

3.4 The Effect of Corporate Governance on Liquidity of Stocks

For any constant number $s_0$ in $[0, \bar{s})$, by equation (3.1), the probability that a noise trader's estimation $\widehat{V}$ on the value of each share relative to its real value is greater than $s_0$ is

$$g_{s_0}(c_m) = 1 - \frac{\rho(c_m) + s_0 - s_0 \cdot \rho(c_m)}{c_m}.$$

Recall that in section 2.1 we clarify that in general-type companies the controlling community will adjust the agency cost $\rho$ as the quality of corporate governance changes. Thus, $\rho$ can be seen as a function of $c_m$ and the form of this function has been given by equation (2.5), i.e.

$$\rho(c_m) = \frac{1-\theta}{2} \cdot f(c_m)$$

In this part, we need to impose a technical and widely used condition to the function f(·). This condition is that f(·) is a concave function of $c_m$, as means that the marginal agency cost extracted by the controlling community is decreasing with the deterioration of corporate governance (i.e. the increase in the upper bound $c_m$ of the company's agency cost).

**Proposition 3**. An increase in the degree of noise traders' participation in trading (i.e. $\lambda$) will enhance liquidity of stocks generally in market. Improvements of corporate governance of the company have the effect of increasing the liquidity of its stocks if $\lambda \neq 0$.

*Proof.*

We use $\frac{\partial K(g,L)}{\partial g}$ and $\frac{\partial K(g,L)}{\partial L}$ to represent the partial derivatives of $K(g, L)$ with respect to its first and second independent variable.

$$\frac{\partial K(g, L)}{\partial L}$$

$$= \sum_{i=1}^{m} \frac{i \cdot L^{i-1}}{i!} e^{-L} - \sum_{i=0}^{m} \frac{L^i}{i!} e^{-L} + \sum_{i=m+1}^{+\infty} \sum_{j=0}^{m} \left[ \frac{iL^{i-1}}{j!\,(i-j)!} e^{-L} (1-g)^j g^{i-j} \right]$$

$$- \sum_{i=m+1}^{+\infty} \sum_{j=0}^{m} \left[ \frac{L^i e^{-L}}{j!\,(i-j)!} (1-g)^j g^{i-j} \right]$$

$$= \sum_{i=1}^{m} \frac{L^{i-1}}{(i-1)!} e^{-L} - \sum_{i=0}^{m} \frac{L^i}{i!} e^{-L} + \sum_{i=m+1}^{+\infty} \sum_{j=0}^{m} \left[ \frac{iL^{i-1}}{j!\,(i-j)!} e^{-L} (1-g)^j g^{i-j} \right]$$

$$- \sum_{i=m+1}^{+\infty} \sum_{j=0}^{m} \left[ \frac{L^i e^{-L}}{j!\,(i-j)!} (1-g)^j g^{i-j} \right]$$

$$= \sum_{i=0}^{m} \frac{L^i}{i!} e^{-L} - \sum_{i=0}^{m} \frac{L^i}{i!} e^{-L}$$

$$+ \sum_{i=m+1}^{+\infty} \sum_{j=0}^{m} \left[ \frac{(i-j)L^{i-1}}{j!\,(i-j)!} e^{-L} (1-g)^j g^{i-j} \right] + \sum_{i=m+1}^{+\infty} \sum_{j=0}^{m} \left[ \frac{jL^{i-1}}{j!\,(i-j)!} e^{-L} (1-g)^j g^{i-j} \right]$$

$$- \left\{ \sum_{i=m}^{+\infty} \sum_{j=0}^{m} \left[ \frac{L^i e^{-L}}{j!\,(i-j)!} (1-g)^j g^{i-j} \right] - \sum_{j=0}^{m} \left[ \frac{L^m e^{-L}}{j!\,(m-j)!} (1-g)^j g^{m-j} \right] \right\}$$

$$= -\frac{L^m}{m!} e^{-L} + \sum_{i=m+1}^{+\infty} \sum_{j=0}^{m} \left[ \frac{L^{i-1}}{j!\,(i-1-j)!} e^{-L} (1-g)^j g^{i-j} \right]$$

$$+ \sum_{i=m+1}^{+\infty} \sum_{j=1}^{m} \left[ \frac{L^{i-1}}{(j-1)!\,(i-j)!} e^{-L} (1-g)^j g^{i-j} \right] - \sum_{i=m}^{+\infty} \sum_{j=0}^{m} \left[ \frac{L^i e^{-L}}{j!\,(i-j)!} (1-g)^j g^{i-j} \right]$$

$$+ \frac{L^m e^{-L}}{m!} \cdot \sum_{j=0}^{m} \left[ \frac{m!}{j!\,(m-j)!} (1-g)^j g^{m-j} \right]$$

$$= -\frac{L^m}{m!} e^{-L} + \sum_{i=m}^{+\infty} \sum_{j=0}^{m} \left[ \frac{L^i}{j!\,(i-j)!} e^{-L} (1-g)^j g^{i-j+1} \right]$$

$$+ \sum_{i=m}^{+\infty} \sum_{j=0}^{m-1} \left[ \frac{L^i e^{-L}}{j!\,(i-j)!} (1-g)^{j+1} g^{i-j} \right] - \sum_{i=m}^{+\infty} \sum_{j=0}^{m} \left[ \frac{L^i e^{-L}}{j!\,(i-j)!} (1-g)^j g^{i-j} \right] + \frac{L^m}{m!} e^{-L}$$

$$= \sum_{i=m}^{+\infty} \sum_{j=0}^{m} \left[ \frac{L^i}{j!\,(i-j)!} e^{-L} (1-g)^j g^{i-j+1} - \frac{L^i e^{-L}}{j!\,(i-j)!} (1-g)^j g^{i-j} \right]$$

$$+ \left\{ \sum_{i=m}^{+\infty} \sum_{j=0}^{m} \left[ \frac{L^i e^{-L}}{j!\,(i-j)!} (1-g)^{j+1} g^{i-j} \right] - \sum_{i=m}^{+\infty} \frac{L^i e^{-L}}{m!\,(i-m)!} (1-g)^{m+1} g^{i-m} \right\}$$

$$= -\sum_{i=m}^{+\infty}\sum_{j=0}^{m}\left[\frac{L^i}{j!\,(i-j)!}e^{-L}(1-g)^j g^{i-j+1}\right] + \sum_{i=m}^{+\infty}\sum_{j=0}^{m}\left[\frac{L^i e^{-L}}{j!\,(i-j)!}(1-g)^{j+1}g^{i-j}\right]$$

$$-\frac{L^m e^{-L}(1-g)^{m+1}}{m!}\cdot\sum_{i=m}^{+\infty}\frac{(gL)^{i-m}}{(i-m)!}$$

$$= -\frac{L^m e^{-L}(1-g)^{m+1}}{m!}\cdot\sum_{j=0}^{+\infty}\frac{(gL)^j}{j!}$$

$$= -\frac{L^m(1-g)^{m+1}}{m!}e^{-L(1-g)} \tag{3.3}$$

Thus, we derive that $\frac{\partial K(g,L)}{\partial L} < 0$. For any constant $s_0$ in $[0,\bar{s})$, we deduce that

$$\frac{\partial ILL_{s_0}(c_m,\lambda)}{\partial \lambda} = -\frac{1}{K(g_{s_0}(c_m),\lambda\Delta T)}\cdot\frac{\partial K(g_{s_0}(c_m),\lambda\Delta T)}{\partial L}\cdot\Delta T > 0.$$

(1) Consider the case where $c_m \leq \rho + s_0(1-\rho)$. Under this case, it's obvious that $\bar{s} = \frac{c_m-\rho}{1-\rho} \leq s_0$, which implies that $F_{s_0}(c_m,\lambda) = 0$ and $ILL_{s_0}(c_m,\lambda) = +\infty$. Therefore, the liquidity of the stocks of the company is highest under this case.

(2) Focus on the case where $c_m > \rho + s_0(1-\rho)$.

$$\frac{\partial K(g,L)}{\partial g} = \sum_{i=m+1}^{+\infty}\sum_{j=0}^{m}\left[\frac{e^{-L}L^i}{j!\,(i-j-1)!}(1-g)^j g^{i-j-1}\right]$$

$$-\sum_{i=m+1}^{+\infty}\sum_{j=1}^{m}\left[\frac{e^{-L}L^i}{(j-1)!\,(i-j)!}(1-g)^{j-1}g^{i-j}\right]$$

$$= \sum_{i=m}^{+\infty}\sum_{j=0}^{m}\left[\frac{e^{-L}L^{i+1}}{j!\,(i-j)!}(1-g)^j g^{i-j}\right] - \sum_{i=m}^{+\infty}\sum_{j=0}^{m-1}\left[\frac{e^{-L}L^{i+1}}{j!\,(i-j)!}(1-g)^j g^{i-j}\right]$$

$$= \sum_{i=m}^{+\infty}\left[\frac{e^{-L}L^{i+1}}{m!\,(i-m)!}(1-g)^m g^{i-m}\right]$$

$$= L^{m+1}(1-g)^m \frac{e^{-L}}{m!}\cdot\sum_{i=m}^{+\infty}\frac{(gL)^{i-m}}{(i-m)!}$$

$$= \frac{L^{m+1}(1-g)^m}{m!}e^{-L(1-g)} \tag{3.4}$$

If $L \neq 0$, then $\frac{\partial K(g,L)}{\partial g} > 0$.

We define $G(c_m) \triangleq \rho(c_m) - c_m\rho'(c_m)$. It's obvious that $G(0) = \rho(0) = 0$. On the other hand, by the concavity of the function $f(\cdot)$, we can deduce that $G'(c_m) = -c_m\rho''(c_m) = -c_m\cdot\frac{1-\theta}{2}\cdot f''(c_m) > 0$. As a result, we derive that

$$G(c_m) > 0.$$

So,

$$\frac{\partial g_{s_0}(c_m)}{\partial c_m} = \frac{(1-s_0)\cdot[\rho(c_m)-c_m\rho'(c_m)] + s_0}{c_m^2} = \frac{(1-s_0)\cdot G(c_m) + s_0}{c_m^2} > 0$$

Based on the above deductions, it follows that

$$\frac{\partial ILL_{s_0}(c_m,\lambda)}{\partial c_m} = -\frac{1}{K(g_{s_0}(c_m),\lambda\Delta T)}\cdot\frac{\partial K(g_{s_0}(c_m),\lambda\Delta T)}{\partial g}\cdot\frac{\partial g_{s_0}(c_m)}{\partial c_m} < 0. \tag{3.5}$$

if $\lambda \neq 0$.

□

3.5 Summarization

Now we explain the economic intuition which underlies the construction of our theoretical model.

For noisy traders who can't precisely know the actual value of the corporate equity by the reasons of their incomplete information about the company, better corporate governance can send them positive signals which reduce the heterogeneity between their beliefs and the extent to which their estimations deviate from the reality as well as have the effect of reducing the probability that they underestimate the value of the equity and strengthening their confidence to invest in the stocks of the company. The noise traders who have received the signals will tend to give a higher bid for each share, thereby providing liquidity for the stocks of the company in the secondary market. The existence of noise traders is an important condition for corporate governance to improve liquidity of the stocks. Based on Proposition 3, we have the following corollary.

**Corollary**. Under the case where there are no noisy traders in the market or they do not participate in any transactions, the effect of corporate governance on the liquidity of the stocks of the company tend to be extremely weak.

*Proof.*

If $\lambda = 0$, for any constant $s_0$ in $[0, \bar{s})$, we can derive

$$\frac{\partial K(g_{s_0}(c_m), \lambda \Delta T)}{\partial g} = \frac{\partial K(g_{s_0}(c_m), 0)}{\partial g} = 0$$

by equation (3.4). Then we conclude that $\frac{\partial ILL_{s_0}(c_m, \lambda)}{\partial c_m} = 0$ by equation (3.5). □

## 4. Synergistic Effect

4.1 Corporate Governance and the Effect of noise trading on liquidity

In proposition 3 we have deduced that an increase in the degree of noise traders' participation in market transactions will promote the liquidity of stocks, as is consistent with Easley, Kiefer and O'hara et al. (1996). We call this phenomenon as the effect of noise traders' providing liquidity for stocks. Consider a company with a parameter $c_m$ of corporate governance. When the degree of noise traders' participation in trading is $\lambda_1$ which satisfies that $\lambda_1 \geq 0$, the liquidity level of its stocks should be $ILL_{s_0}(c_m, \lambda_1)$. If this degree then increases to $\lambda_2$ which satisfies that $\lambda_2 > \lambda_1$, the liquidity level will be lifted to $ILL_{s_0}(c_m, \lambda_2)$. The increase is

$$\Delta_\lambda ILL_{s_0}(c_m) \triangleq ILL_{s_0}(c_m, \lambda_2) - ILL_{s_0}(c_m, \lambda_1) > 0.$$

$\Delta_\lambda ILL_{s_0}(c_m)$ essentially quantifies the effect of noise traders' providing liquidity for the stocks of this company. What we are interested in is whether $\Delta_\lambda ILL_{s_0}(c_m)$ changes with the quality of corporate governance $c_m$.

Suppose that there are two listed companies in the market and their parameters of corporate governance are respectively $c_m^g$ and $c_m^b$ which satisfy that $c_m^g < c_m^b$. Additionally, the other parameters of them are almost the same. Based on the framework of the theoretical model in Section 3, we draw the following conclusion:

**Proposition 4**. For companies with better corporate governance, the effect of noise traders' providing liquidity for their stocks is more intense. In other words, when more noise traders begin to participate in market trading, the liquidity level of the stocks of these companies will increase more greatly compared to the companies with worse corporate governance, i.e. $\Delta_\lambda ILL_{s_0}(c_m^g) > \Delta_\lambda ILL_{s_0}(c_m^b)$.

*Proof.*

Since

$$\Delta_\lambda ILL_{s_0}(c_m) = ILL_{s_0}(c_m, \lambda_2) - ILL_{s_0}(c_m, \lambda_1) = \int_{\lambda_1}^{\lambda_2} \frac{\partial ILL_{s_0}(c_m, \lambda)}{\partial \lambda} d\lambda,$$

in order to prove that $\Delta_\lambda ILL_s(c_m^g) > \Delta_\lambda ILL_s(c_m^b)$, we only need to demonstrate that

$$\frac{\partial ILL_{s_0}(c_m^g, \lambda)}{\partial \lambda} > \frac{\partial ILL_{s_0}(c_m^b, \lambda)}{\partial \lambda} \tag{4.1}$$

holds for any positive number $\lambda$. We define $H(g, L) \triangleq -K(g, L)\big/\frac{\partial K(g,L)}{\partial L}$, then

$$\frac{\partial ILL_{s_0}(c_m, \lambda)}{\partial \lambda} = -\frac{1}{K(g_{s_0}(c_m), \lambda \Delta T)} \cdot \frac{\partial K(g_{s_0}(c_m), \lambda \Delta T)}{\partial L} \cdot \Delta T$$

$$= \frac{\Delta T}{H(g_{s_0}(c_m), \lambda \Delta T)}$$

Let $\frac{\partial H(g,L)}{\partial g}$ denote the partial derivative of $H(g, L)$ with respect to its first independent variable, then

$$\frac{\partial^2 ILL_{s_0}(c_m, \lambda)}{\partial c_m \partial \lambda} = -\frac{\Delta T}{H^2(g_{s_0}(c_m), \lambda \Delta T)} \cdot \frac{\partial H(g_{s_0}(c_m), \lambda \Delta T)}{\partial g} \cdot \frac{\partial g_{s_0}(c_m)}{\partial c_m} \quad (4.2)$$

We have proved that $\partial g_{s_0}(c_m)/\partial c_m > 0$ in Proposition 3.

On the other hand,

$$\frac{\partial H(g, L)}{\partial g} = \left(\frac{\partial K(g, L)}{\partial L}\right)^{-2} \left(K(g, L) \frac{\partial^2 K(g, L)}{\partial g \partial L} - \frac{\partial K(g, L)}{\partial L} \frac{\partial K(g, L)}{\partial g}\right).$$

By equation (3.3), we can deduce that

$$\frac{\partial^2 K(g, L)}{\partial g \partial L} = [(m+1) - L(1-g)] \cdot \frac{L^m(1-g)^m e^{-L(1-g)}}{m!}.$$

$$K(g, L) \frac{\partial^2 K(g, L)}{\partial g \partial L} - \frac{\partial K(g, L)}{\partial L} \frac{\partial K(g, L)}{\partial g}$$

$$= \left\{\sum_{i=0}^{m} \frac{L^i}{i!} e^{-L} + \sum_{i=m+1}^{+\infty} \sum_{j=0}^{m} \left[\frac{L^i e^{-L}}{j!(i-j)!}(1-g)^j g^{i-j}\right]\right\}[(m+1) - L(1-g)]$$

$$\cdot \frac{L^m(1-g)^m e^{-L(1-g)}}{m!} + \frac{L^{2m+1}(1-g)^{2m+1}}{(m!)^2} e^{-2L(1-g)}$$

$$= \frac{L^m(1-g)^m e^{gL-2L}}{m!} \left\{\sum_{i=0}^{m} \frac{L^i}{i!}[(m+1-i) + i - L(1-g)]\right.$$

$$+ \sum_{i=m+1}^{+\infty} \sum_{j=0}^{m} \frac{L^i}{j!(i-j)!}(1-g)^j g^{i-j}[(m+1-j) + j - L(1-g)]$$

$$\left. + \frac{L^{m+1}(1-g)^{m+1}}{m!} e^{gL}\right\}$$

$$= \frac{L^m(1-g)^m e^{gL-2L}}{m!} \cdot \left\{\sum_{i=0}^{m} \frac{L^i}{i!}(m+1-i) + \sum_{i=1}^{m} \frac{L^i}{(i-1)!} - \sum_{i=0}^{m} \frac{L^{i+1}}{i!}(1-g)\right.$$

$$+ \sum_{i=m+1}^{+\infty} \sum_{j=0}^{m} \frac{L^i}{j!(i-j)!}(1-g)^j g^{i-j}(m+1-j) + \sum_{i=m+1}^{+\infty} \sum_{j=1}^{m} \frac{L^i(1-g)^j g^{i-j}}{(j-1)!(i-j)!}$$

$$\left. - \sum_{i=m+1}^{+\infty} \sum_{j=0}^{m} \frac{L^{i+1}}{j!(i-j)!}(1-g)^{j+1} g^{i-j} + \frac{L^{m+1}(1-g)^{m+1}}{m!} e^{gL}\right\}$$

$$> \frac{L^m(1-g)^m e^{gL-2L}}{m!} \left\{\sum_{i=1}^{m} \frac{L^i}{(i-1)!} - \sum_{i=0}^{m} \frac{L^{i+1}}{i!}(1-g) + \sum_{i=m+1}^{+\infty} \sum_{j=1}^{m} \frac{L^i(1-g)^j g^{i-j}}{(j-1)!(i-j)!}\right.$$

$$\left. - \sum_{i=m+1}^{+\infty} \sum_{j=0}^{m} \frac{L^{i+1}}{j!(i-j)!}(1-g)^{j+1} g^{i-j} + \frac{L^{m+1}(1-g)^{m+1}}{m!} e^{gL}\right\}$$

$$= \frac{L^m(1-g)^m e^{gL-2L}}{m!} \left\{\sum_{i=0}^{m-1} \frac{L^{i+1}}{i!} - \sum_{i=0}^{m} \frac{L^{i+1}}{i!}(1-g) + \sum_{i=m}^{+\infty} \sum_{j=0}^{m-1} \frac{L^{i+1}(1-g)^{j+1} g^{i-j}}{j!(i-j)!}\right.$$

$$
\begin{aligned}
&\quad \left. - \sum_{i=m+1}^{+\infty} \sum_{j=0}^{m} \frac{L^{i+1}}{j!\,(i-j)!} (1-g)^{j+1} g^{i-j} + \frac{L^{m+1}(1-g)^{m+1}}{m!} e^{gL} \right\} \\
&= \frac{L^m (1-g)^m e^{gL-2L}}{m!} \left\{ \sum_{i=0}^{m-1} \frac{L^{i+1}}{i!} \cdot g - \frac{L^{m+1}}{m!}(1-g) + \sum_{i=m}^{+\infty}\sum_{j=0}^{m} \frac{L^{i+1}(1-g)^{j+1} g^{i-j}}{j!\,(i-j)!} \right. \\
&\qquad\qquad \left. - \sum_{i=m}^{+\infty} \frac{L^{i+1}(1-g)^{m+1} g^{i-m}}{m!\,(i-m)!} - \sum_{i=m+1}^{+\infty}\sum_{j=0}^{m} \frac{L^{i+1}}{j!\,(i-j)!}(1-g)^{j+1} g^{i-j} \right. \\
&\qquad\qquad \left. + \frac{L^{m+1}(1-g)^{m+1}}{m!} e^{gL} \right\} \\
&> \frac{L^m (1-g)^m e^{gL-2L}}{m!} \left\{ -\frac{L^{m+1}}{m!}(1-g) + \sum_{j=0}^{m} \frac{L^{m+1}(1-g)^{j+1} g^{m-j}}{j!\,(m-j)!} \right. \\
&\qquad\qquad \left. - \sum_{i=m}^{+\infty} \frac{L^{i+1}(1-g)^{m+1} g^{i-m}}{m!\,(i-m)!} + \frac{L^{m+1}(1-g)^{m+1}}{m!} e^{gL} \right\} \\
&= \frac{L^m (1-g)^m e^{gL-2L}}{m!} \left\{ -\frac{L^{m+1}}{m!}(1-g) + \frac{L^{m+1}(1-g)}{m!} \sum_{j=0}^{m} \binom{m}{j}(1-g)^j g^{m-j} \right. \\
&\qquad\qquad \left. - \frac{L^{m+1}(1-g)^{m+1}}{m!} \sum_{i=m}^{+\infty} \frac{(gL)^{i-m}}{(i-m)!} + \frac{L^{m+1}(1-g)^{m+1}}{m!} e^{gL} \right\} \\
&= \frac{L^m (1-g)^m e^{gL-2L}}{m!} \left\{ -\frac{L^{m+1}}{m!}(1-g) + \frac{L^{m+1}(1-g)}{m!} - \frac{L^{m+1}(1-g)^{m+1}}{m!} \sum_{k=0}^{+\infty} \frac{(gL)^k}{k!} \right. \\
&\qquad\qquad \left. + \frac{L^{m+1}(1-g)^{m+1}}{m!} e^{gL} \right\} \\
&= \frac{L^m (1-g)^m e^{gL-2L}}{m!} \left\{ -\frac{L^{m+1}(1-g)^{m+1}}{m!} e^{gL} + \frac{L^{m+1}(1-g)^{m+1}}{m!} e^{gL} \right\} \\
&= 0.
\end{aligned}
$$

Therefore, we have derived that $\frac{\partial H(g,L)}{\partial g} > 0$. We can further infer that $\frac{\partial^2 \mathrm{ILL}_{s_0}(c_m,\lambda)}{\partial c_m \partial \lambda} < 0$ by equation (4.2) and it follows that equation (4.1) holds. The proof of this proposition is completed. □

Based on Proposition 4, we can infer that when the market trend is upward and various investors become more active in investment, the liquidity of the stocks of the companies with better corporate governance will increase more than those with worse corporate governance. On the other hand, when the market trend is downward and the activity of various investors in investment is weakened, the liquidity of the stocks of the companies with better corporate governance may decrease more than those with worse corporate governance, but by Proposition 3 we can affirm that the liquidity level of the former is still higher than the latter under this situation.

4.2 Noise Trading and the Effect of Corporate Governance on liquidity

In proposition 3 we have deduced that an improvement in corporate governance of a company will promote the liquidity of its stocks, as is consistent with the empirical evidence of Chung, Elder, Kim(2010). We call this phenomenon as the effect of corporate governance on liquidity of stocks. Suppose that the degree of noise traders' participation in trading is $\lambda$. For a representative company with a parameter $c_m^1$ of corporate governance, the liquidity level of its stocks should be $\mathrm{ILL}_{s_0}(c_m^1, \lambda)$ at this time. If the corporate governance in the company is improved and the parameter of corporate governance consequently decreases to $c_m^2$ ($c_m^2 < c_m^1$), the liquidity level will be lifted to $\mathrm{ILL}_{s_0}(c_m^2, \lambda)$. The increase is

$$\Delta_c \text{ILL}_{s_0}(\lambda) \triangleq \text{ILL}_{s_0}(c_m^2, \lambda) - \text{ILL}_{s_0}(c_m^1, \lambda) > 0.$$

$\Delta_c \text{ILL}_{s_0}(\lambda)$ essentially quantifies the effect of corporate governance on the liquidity of its stocks. What we are interested in is whether $\Delta_c \text{ILL}_{s_0}(\lambda)$ changes with the degree of noise traders' participation in trading $\lambda$.

Suppose that there are two stages of the market. In the first stage the degree of noise traders' participation in trading is $\lambda^l$ and in the second stage this degree rises to $\lambda^h$ (i.e. $\lambda^l < \lambda^h$). The company can improve its corporate governance and reduce the parameter from $c_m^1$ to $c_m^2$ in one of the two stages. Based on the framework of the theoretical model in Section 3, we draw the following conclusion:

**Proposition 5**. The effect of corporate governance on the liquidity of the company's stocks will become more intense when more noise traders participate in market trading. In other words, improving corporate governance in the stage when more noise traders participate in trading can increase the liquidity of the company's stocks more greatly compared to the stage when less noise traders participate, i.e. $\Delta_c \text{ILL}_{s_0}(\lambda^h) > \Delta_c \text{ILL}_{s_0}(\lambda^l)$.

*Proof.*

Since

$$\Delta_c \text{ILL}_{s_0}(\lambda) = \text{ILL}_{s_0}(c_m^2, \lambda) - \text{ILL}_{s_0}(c_m^1, \lambda) = \int_{c_m^2}^{c_m^1} - \frac{\partial \text{ILL}_{s_0}(c_m, \lambda)}{\partial c_m} d\lambda,$$

in order to prove that $\Delta_c \text{ILL}_{s_0}(\lambda^h) > \Delta_c \text{ILL}_{s_0}(\lambda^l)$, we only need to demonstrate that

$$\frac{\partial \text{ILL}_{s_0}(c_m, \lambda^h)}{\partial c_m} < \frac{\partial \text{ILL}_{s_0}(c_m, \lambda^l)}{\partial c_m} \quad (4.3)$$

holds for any positive number $\lambda$. We define $J(g, L) \triangleq - K(g, L) \Big/ \frac{\partial K(g,L)}{\partial g}$, then

$$\frac{\partial \text{ILL}_{s_0}(c_m, \lambda)}{\partial c_m} = -\frac{1}{K(g_{s_0}(c_m), \lambda \Delta T)} \cdot \frac{\partial K(g_{s_0}(c_m), \lambda \Delta T)}{\partial g} \cdot \frac{\partial g_{s_0}(c_m)}{\partial c_m}$$

$$= \frac{1}{J(g_{s_0}(c_m), \lambda \Delta T)} \cdot \frac{\partial g_{s_0}(c_m)}{\partial c_m}.$$

Let $\frac{\partial J(g,L)}{\partial L}$ denote the partial derivative of $J(g, L)$ with respect to its second independent variable, then

$$\frac{\partial^2 \text{ILL}_{s_0}(c_m^g, \lambda)}{\partial \lambda \partial c_m} = -\frac{\Delta T}{J^2(g_{s_0}(c_m), \lambda \Delta T)} \cdot \frac{\partial J(g_{s_0}(c_m), \lambda \Delta T)}{\partial L} \cdot \frac{\partial g_{s_0}(c_m)}{\partial c_m} \quad (4.4)$$

It has been proved that $\partial g_{s_0}(c_m)/\partial c_m > 0$ in Proposition 3.

On the other hand,

$$\frac{\partial J(g, L)}{\partial L} = \left(\frac{\partial K(g, L)}{\partial g}\right)^{-2} \left(K(g, L) \frac{\partial^2 K(g, L)}{\partial L \partial g} - \frac{\partial K(g, L)}{\partial L} \frac{\partial K(g, L)}{\partial g}\right).$$

Using equation (3.3) and (3.4), it is easy to verify that

$$\frac{\partial^2 K(g, L)}{\partial L \partial g} = \frac{\partial^2 K(g, L)}{\partial g \partial L}.$$

It has been shown that

$$K(g, L) \frac{\partial^2 K(g, L)}{\partial g \partial L} - \frac{\partial K(g, L)}{\partial L} \frac{\partial K(g, L)}{\partial g} > 0,$$

in Proposition 4. Therefore,

$$K(g, L) \frac{\partial^2 K(g, L)}{\partial L \partial g} - \frac{\partial K(g, L)}{\partial L} \frac{\partial K(g, L)}{\partial g} > 0.$$

As a result, $\frac{\partial J(g,L)}{\partial L} > 0$ and we can further derive that $\frac{\partial^2 \text{ILL}_{s_0}(c_m^g, \lambda)}{\partial \lambda \partial c_m} < 0$ by equation (4.4). It follows that equation (4.3) holds. The proof of this proposition is completed.

□

Combining Proposition 4 and Proposition 5, we can conclude that corporate governance and the degree of noise traders' participation in trading have a synergistic effect on increasing the liquidity of a company's stocks.

## 5. Conclusions for Controlled-type Companies

First of all, we suppose that for any listed company in market, noise traders don't know whether it is a general-type or controlled-type company.

The conclusions in section 3 and 4 are for general-type companies, while in this section we focus on whether these conclusions is applicable for controlled-type companies which are special.

The company of this type tends to have a relatively fixed agency cost the size of which is denoted by $\rho_0$, and its parameter $c_m$ of corporate governance is also largely determined by its controlling community. The controlling community has tight control over the company, as causes that the probability of their expropriation being revealed (i.e. $p(c_m)$) and the punishment for their expropriation (i.e. $d(c_m)$) tend to be negligible. Therefore, the agency cost $\rho$ of the company will not be determined by the quality of corporate governance in the way we have elaborated in section 2.1. The parameter $c_m$ which is determined by the controlling community should satisfies $c_m \geq \rho_0$ to ensure that their expropriation can't be restricted by corporate governance. For the company of this kind, we treat its agency cost $\rho_0$ as an exogenous parameter which is determined by the history and life-cycle of the company. However, it's easy to verify that the agency cost can still affect the outside investors' of its stocks in the way we have analyzed in section 2.2, that is, the formulas (2.8) and (2.9) are still valid for the company.

For the company in this kind, by formula (3.1), the probability that the discount rate of a noise trader's estimation $\widehat{V}$ on each share relative to its real value $V(\rho)$ is larger than $s_0$ ($s_0 \in [0, \overline{s})$) should be

$$g_{s_0}(c_m) = 1 - \frac{\rho_0 + s_0 - s_0 \cdot \rho_0}{c_m}, \quad (5.1)$$

for any $s_0 \in [\overline{s}, 1)$, $g_{s_0}(c_m) = 0$. Here, we deem the agency cost $\rho_0$ constant. Though $\rho_0$ satisfies that $\rho_0 < c_m$, it doesn't change with the parameter $c_m$ of corporate governance, as differs from general-type companies. We have the following conclusions for controlled-type companies.

**Proposition 6**. Suppose that $M \geq N_I$. At the time of $T_2$, for any $s_0 \in [0, \overline{s})$, the probability that the discount rate of the market price of each share relative to its fair value is larger than $s_0$ is:

$$F_{s_0}(c_m, \lambda) = K\big(g_{s_0}(c_m), K\big(g_{s_0}(c_m), \lambda \Delta T\big)\big),$$

where $K(\cdot,\cdot)$ and $g_{s_0}(\cdot)$ are defined by equation (3.2) and (5.1) respectively. Additionally, for any $s_0 \in [\overline{s}, 1)$, $F_{s_0}(c_m, \lambda) = 0$.

□

The proof of Proposition 6 is almost identical to that of Proposition 2. Based on Proposition 6, we can derive the following proposition which is similar to Proposition 3.

**Proposition 7**. An increase in the degree of noise traders' participation in trading (i.e. $\lambda$) will enhance liquidity of stocks generally in market. Improvements of corporate governance of the company have the effect of increasing the liquidity of its stocks if $\lambda \neq 0$.

*Proof.*

The process of its proof is almost the same as that of Proposition 3. We only need to notice that it still holds that $g_{s_0}{'}(c_m) < 0$ for any $s_0 \in [0, \overline{s})$, though there is some difference between $g_{s_0}(\cdot)$ defined by equation (5.1) and that defined by equation (3.1).

□

Proposition 7 indicates that for the company of this special type, corporate governance still has the effect to increase the liquidity of its stocks, though the constraint of corporate governance on the scale of its agency cost is very weak and its agency cost tend to be fixed. Simply speaking, the reason is that better corporate governance can send positive signals to noise traders who have incomplete information and need to estimate the value of its stocks, as reduces the probability that they underestimate the value and motivates them to provide liquidity for its stocks with a higher bid. This is also the economic intuition that underlies our deduction in this part.

In addition, the conclusions of Proposition 4 and Proposition 5 also hold for the companies of this special type and the proofs of them are almost identical to those of Proposition 4 and Proposition 5, so we will not repeat them.

## 6. Conclusions and Suggestions

Summarizing the analyses and derivations in this paper, we conclude that good corporate governance has the effect of increasing liquidity of stocks of listed companies, but an important condition for apparent occurrence of this effect is that there exists noise traders with incomplete information in the market and they are involved in trading to some extent. Our theoretical model further implies that an increase in the degree of noise traders' participation in trading can enhancing this effect. Additionally, it has been widely accepted that an increase in the degree of noise traders' participating in trading has the effect of promoting the liquidity of stocks. Our model also implies that an improvement in corporate governance can reinforcing this effect. Thus, we reveal that corporate governance and the degree of noise traders' participation have a synergistic effect in elevating the liquidity level of stocks.

For investors, buying and holding the stocks of the companies with good corporate governance is a potential channel to improve the liquidity of their portfolios, especially during the periods when noise traders are active in investment. For those investors who have to sell out a large number of stocks shortly, it may be sensible for them to give priority to selling the stocks of companies with relatively better corporate governance. This strategy is likely to reduce their loss caused by the large sales volume in a short time. For a listed company, improving corporate governance is a potential channel to increase the liquidity of its stocks in the secondary market, especially during the periods when noise traders are active in investment.